\documentclass[twoside,twocolumn,9pt]{article}
\usepackage{extsizes}
\usepackage[super,sort&compress,comma]{natbib} 
\usepackage[version=3]{mhchem}
\usepackage[left=1.5cm, right=1.5cm, top=1.785cm, bottom=2.0cm]{geometry}
\usepackage{balance}
\usepackage{times,mathptmx}
\usepackage{sectsty}
\usepackage{graphicx} 
\usepackage{lastpage}
\usepackage[format=plain,justification=justified,singlelinecheck=false,font={stretch=1.125,small,sf},labelfont=bf,labelsep=space]{caption}
\usepackage{float}
\usepackage{fancyhdr}
\usepackage{fnpos}
\usepackage[english]{babel}
\addto{\captionsenglish}{
  
}
\usepackage{array}
\usepackage{droidsans}
\usepackage{charter}
\usepackage[T1]{fontenc}
\usepackage[usenames,dvipsnames]{xcolor}
\usepackage{setspace}
\usepackage[compact]{titlesec}
\usepackage{hyperref}
\usepackage[english]{babel} 

\newcommand{\cme}{cm$^{-1}$}
\newcommand{\ST}{S$_1\leftarrow$S$_0$}
\newcommand{\mz}{$m/z$}
\newcommand{\npoo}{NPOO$^-$}
\newcommand{\npoc}{NPOC$^-$}

\usepackage{epstopdf}

\definecolor{cream}{RGB}{222,217,201}

\begin{document}

\pagestyle{fancy}
\thispagestyle{plain}
\fancypagestyle{plain}{

\renewcommand{\headrulewidth}{0pt}
}

\makeFNbottom
\makeatletter
\renewcommand\LARGE{\@setfontsize\LARGE{15pt}{17}}
\renewcommand\Large{\@setfontsize\Large{12pt}{14}}
\renewcommand\large{\@setfontsize\large{10pt}{12}}
\renewcommand\footnotesize{\@setfontsize\footnotesize{7pt}{10}}
\makeatother

\renewcommand{\thefootnote}{\fnsymbol{footnote}}
\renewcommand\footnoterule{\vspace*{1pt}%
\color{cream}\hrule width 3.5in height 0.4pt \color{black}\vspace*{5pt}} 
\setcounter{secnumdepth}{5}

\makeatletter 
\renewcommand\@biblabel[1]{#1}            
\renewcommand\@makefntext[1]%
{\noindent\makebox[0pt][r]{\@thefnmark\,}#1}
\makeatother 
\renewcommand{\figurename}{\small{Fig.}~}
\sectionfont{\sffamily\Large}
\subsectionfont{\normalsize}
\subsubsectionfont{\bf}
\setstretch{1.125} 
\setlength{\skip\footins}{0.8cm}
\setlength{\footnotesep}{0.25cm}
\setlength{\jot}{10pt}
\titlespacing*{\section}{0pt}{4pt}{4pt}
\titlespacing*{\subsection}{0pt}{15pt}{1pt}

\fancyfoot{}
\fancyfoot[RO]{\footnotesize{\sffamily{1--\pageref{LastPage} ~\textbar  \hspace{2pt}\thepage}}}
\fancyfoot[LE]{\footnotesize{\sffamily{\thepage~\textbar\hspace{3.45cm} 1--\pageref{LastPage}}}}
\fancyhead{}
\renewcommand{\headrulewidth}{0pt} 
\renewcommand{\footrulewidth}{0pt}
\setlength{\arrayrulewidth}{1pt}
\setlength{\columnsep}{6.5mm}
\setlength\bibsep{1pt}

\makeatletter 
\newlength{\figrulesep} 
\setlength{\figrulesep}{0.5\textfloatsep} 

\newcommand{\topfigrule}{\vspace*{-1pt}%
\noindent{\color{cream}\rule[-\figrulesep]{\columnwidth}{1.5pt}} }

\newcommand{\botfigrule}{\vspace*{-2pt}%
\noindent{\color{cream}\rule[\figrulesep]{\columnwidth}{1.5pt}} }

\newcommand{\dblfigrule}{\vspace*{-1pt}%
\noindent{\color{cream}\rule[-\figrulesep]{\textwidth}{1.5pt}} }

\makeatother

\twocolumn[
  \begin{@twocolumnfalse}
\vspace{3cm}
\sffamily
\begin{tabular}{m{4.5cm} p{13.5cm} }

 & \noindent\LARGE{\textbf{Photodetachment and Photoreactions of Substituted Naphthalene Anions in a Tandem Ion Mobility Spectrometer$^\dag$}} \\
\vspace{0.3cm} & \vspace{0.3cm} \\

 & \noindent\large{James N. Bull,\textit{$^{a}$} Jack T. Buntine,\textit{$^{a}$} Michael S. Scholz,\textit{$^{a}$} Eduardo Carrascosa,\textit{$^{a}$} Linda Giacomozzi,\textit{$^{b}$}  Mark H. Stockett,\textit{$^{b}$} and Evan J. Bieske\textit{$^{a}$}}$^{\ast}$\ \\

 & \noindent\normalsize{Substituted naphthalene anions (deprotonated 2-naphthol and 6-hydroxy-2-naphthoic acid) are spectroscopically probed in a tandem drift tube ion mobility spectrometer (IMS). 
Target anions are selected according to their drift speed through nitrogen buffer gas in the first IMS stage before being exposed to a pulse of tunable light that induces either photodissociation or electron photodetachment, which is conveniently monitored by scavenging the detached electrons with trace \ce{SF6} in the buffer gas.  The photodetachment action spectrum of the 2-naphtholate anion exhibits a band system spanning 380-460\,nm~with a prominent series of peaks spaced by 440\,\cme, commencing at 458.5\,nm, and a set of weaker peaks near the electron detachment threshold corresponding to transitions to dipole-bound states. The two deprotomers of 6-hydroxy-2-naphthoic acid are separated and spectroscopically probed independently. The molecular anion formed from deprotonation of the hydroxy group possesses a photodetachment action spectrum similar to that of the 2-naphtholate anion with an onset at 470\,nm and a maximum at 420\,nm. Near threshold,  photoreaction with \ce{SF6} is observed with displacement of an OH group by an F atom. In contrast, the anion formed from deprotonation of the carboxylic acid group  features a photodissociation action spectrum, recorded on the \ce{CO2} loss channel, lying to much shorter wavelength with an onset at 360\,nm and maximum photoresponse at 325\,nm.  } \\

\end{tabular}

 \end{@twocolumnfalse} \vspace{0.6cm}

  ]

\renewcommand*\rmdefault{bch}\normalfont\upshape
\rmfamily
\section*{}
\vspace{-1cm}


\footnotetext{\textit{$^{a}$~School of Chemistry, University of Melbourne, Australia. Tel: +61 383447082; E-mail: evanjb@unimelb.edu.au}}
\footnotetext{\textit{$^{b}$~Department of Physics, Stockholm University, Sweden. }}

\footnotetext{\dag~Electronic Supplementary Information (ESI) available: Synthesis procedure for 6-hydroxy- 2-methylnaphthoate, calculated vibrational frequencies for the naphtholate anion, photodetachment action spectra of 2-naphtholate anion, 1-naphtholate anion, the two deprotomers of 6-hydroxy-2-naphthoic acid, and deprotonated 6-hydroxy- 2-methylnaphthoate. See DOI: 10.1039/b000000x/}



\section{Introduction}
Polycyclic Aromatic Hydrocarbons (PAHs) are postulated to be ubiquitous in the interstellar medium (ISM) \cite{tielens08}, not only as free gas-phase molecules but also condensed into the icy mantles of dust grains \cite{Hardegree-Ullman2014}, meteorites and other interplanetary particles \cite{Allamandola1987,Clemett1993}. 
Pioneering experimental studies have shown that irradiation of PAHs in interstellar ice analogues by ultraviolet light or high-energy particles gives rise to the addition of alcohol (--OH), quinone (=O), and carboxylic acid (--COOH) groups, among others \cite{Bernstein1999,Bernstein2002,Bernstein2002a,Cook2015}. It has been suggested that energetic processing of frozen PAHs contributes to the diversity of the interstellar organic inventory.\cite{tielens08}
In this contribution, we report the results of photochemical and spectroscopic experiments on several substituted naphthalene anions with deprotonated alcohol and carboxylic side groups. Such species could be released into the ISM after being formed in interstellar ices. Our goal is to better understand their gas-phase photochemistry and stability to help assess their possible role in astrochemistry.

\begin{figure}[h]
\centering
  \includegraphics[width=6.5cm]{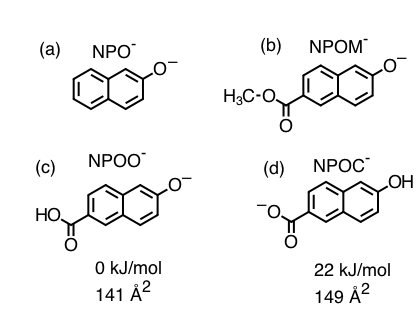}
  \caption{Target molecular anions: (a) 2-naphtholate anion (NPO$^-$), (b) methyl ester of 6-carboxy-2-naphtholate anion  (NPOM$^-$), (c) 6-carboxy-2-naphtholate anion (\npoo), and (d) 6-hydroxy-2-naphthoate anion (\npoc).}
    \label{f1}
\end{figure}

The electronic transitions of deprotonated 2-naphthol and 6-hydroxy-2-naphthoic acid are investigated using a tandem ion mobility spectrometer (IMS) coupled to a tunable-wavelength laser in an IMS-laser-IMS configuration. The target molecular anions are shown in Fig.\,\ref{f1}. 
To our knowledge, there is limited information on the intrinsic electronic absorptions of naphtholate anions in the gas phase, aside from recent photoelectron studies of 1-naphtholate and 2-naphtholate anions aimed mainly at understanding the properties of the 1-naphthoxy and 2-naphthoxy radicals, which focused on transitions to dipole-bound states near the electron detachment threshold.\cite{Kregel:2018dc} The 6-hydroxy-2-naphthoic acid molecule possesses two deprotonation sites with the resulting deprotomers (Fig.\,\ref{f1}(c) and (d)) expected to have quite different electronic properties. In the gas phase, the naphtholate form (Fig.\,\ref{f1}(c), \npoo) is predicted to lie 22\,kJ/mol lower in energy than the carboxylate form (Fig.\,\ref{f1}(d), \npoc), whereas in solution  the carboxylate form is favoured. Using the IMS-laser-IMS approach it is possible to independently probe the electronic transitions of the naphtholate and carboxylate forms and to follow laser-induced reactions.

In general, characterising the properties of molecular ions in the gas phase is an essential precursor to understanding their behaviour in condensed phases where complications inevitably arise due to micro-environmental interactions. Probing molecular ions in the gas phase using standard absorption techniques is often difficult due to low ion densities and the presence of many different absorbing species. On the positive side, it is straightforward to guide, mass-select, and trap molecular ions using electric or magnetic fields, and to detect single ions. Therefore,  spectroscopic approaches have commonly relied upon combinations of mass spectrometers and  tunable lasers to promote photofragmentation or photodetachment (for anions). In situations where the photon energy is insufficient to break a molecular bond, the target molecular ion can be tagged with a rare gas atom that is dislodged following resonant excitation. This strategy has been used to obtain both electronic and infrared spectra of an enormous range of molecular ions.\cite{Duncan:2000dp} 

Isomer-selective approaches have been developed to probe molecules that exist in several different forms. For example, hole burning strategies can be used to distinguish isomer populations for charged molecules in the gas phase.\cite{Rizzo:2009ty}\cite{Wolk:2014fy} For neutral molecules, Stark deflection has been used to separate isomers and conformers prior to spectroscopic interrogation.\cite{Filsinger:2009ij, Chang:2013dm} Another approach for selecting molecular ions involves using a drift tube ion mobility spectrometer (IMS), whereby charged molecules are separated according to their collision cross-sections as they are propelled by an electric field through a neutral buffer gas (usually He or \ce{N2}). Folded, compact ions travel more quickly than unfolded, extended ions. IMS has become a widespread approach for separating and characterising molecular isomers, particularly proteins and biomolecules.\cite{Pringle:2007ib}  Recently, drift-tube ion mobility spectrometers have been used to separate charged molecular isomers prior to laser excitation and photofragmentation. For example, this scheme was used to obtain infrared multiphoton dissociation spectra of the two forms of protonated benzocaine, in which the proton  resides either on the \ce{NH2} group or the carbonyl oxygen.\cite{Warnke:2015kw}  A similar scheme, whereby isomers are mobility-selected prior to confinement in a cryogenic ion trap, has been deployed to obtain spectra of polypeptides and other biomolecules. \cite{Masson:2015fe} In other studies, isomer anions have been separated in a drift tube IMS prior to being probed through photoelectron spectroscopy.\cite{Vonderach:2011dk}

An alternative approach is to select the target isomer ions in a first IMS stage, excite them with a laser pulse, separate the photoisomers in a second IMS stage, with a final mass spectrometer stage providing some surety that the selected isomer indeed has the expected mass. By monitoring the photoisomer signal as a function of laser wavelength one can generate a photoisomerisation action spectrum (PISA spectrum). The use of tandem IMS-IMS to monitor collision induced isomerisation of mobility-selected molecular ions was pioneered by Clemmer and coworkers,\cite{Koeniger:2006hm} and has been extended to follow photoinduced conformational changes in a range of different molecules in the gas phase, including retinal protonated Schiff base, flavins, polyenes and merocyanines.\cite{Adamson:2013kg}\cite{Simon:2015eq}\cite{Bull2018} There are several advantages of the strategy, including isomer specificity, and the ability to monitor photoisomerisation, photodissociation and photodetachment (for anions). One current limitation is that the drift tube is at 300\,K so that spectral features tend to be broad due to contributions from  vibrational hot bands and extended rotational band profiles. In the future, narrower, more informative action spectra may be obtained using drift tubes that are cooled in the section in which the ions are intercepted by the laser beam. In addition, cooled drift tubes offer enhanced mobility resolution and a capacity to distinguish ions with similar collision cross-sections.\cite{Servage:2016ci}

\section{Experimental approach}
\begin{figure}
 \centering
 \includegraphics[width=8.5cm]{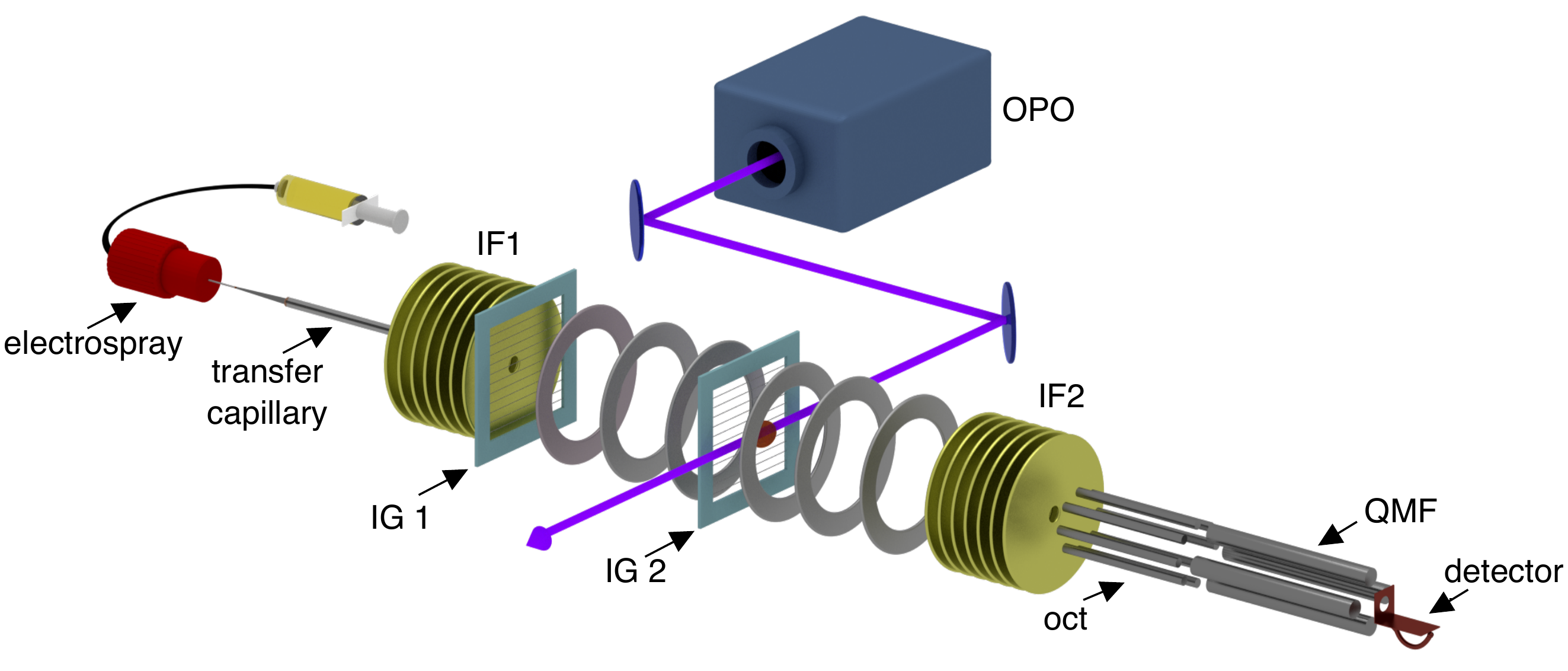}
 \caption{Tandem ion mobility spectrometer. Ions produced by an electrospray ion source are collected by an ion funnel (IF1), pulsed into a drift tube through an electrostatic ion gate (IG1), and propelled through \ce{N2} buffer gas (6 Torr) by an electric field established by applying potentials to a series of ring electrodes. Approximately half way along the drift region the ion packets can be gated by an electrostatic ion gate (IG2), shortly after which they are intercepted by a pulse of light from a tunable optical parametric oscillator (OPO). Photoisomers and photofragments are separated from the parent ions in the second stage of the drift region and quadrupole mass filter (QMF) before striking the ion detector.}
 \label{f2}
\end{figure}

The electronic spectra of the substituted naphthalenes were obtained either through resonance enhanced photodetachment or through resonance enhanced photodissociation. 
A schematic view of the experimental arrangement is shown in Fig.\,\ref{f2}. Briefly, deprotonated anions were produced using an electrospray ionization (ESI) source coupled through a heated desolvation capillary to a tandem ion mobility spectrometer. After passing through the capillary the ions were collected radially by an ion funnel, and injected through a pulsed electrostatic ion gate (IG1 - opening time 100\,$\mu$s) into the drift region where they were propelled through \ce{N2} buffer gas (P$\approx$6\,Torr) by a 44\,V/cm electric field sustained by a series of ring electrodes. After drifting for $\approx$50\,cm, the ions encountered an electrostatic Bradbury-Nielsen ion gate (IG2) that could be opened to select ions with desired mobility. Immediately after the ion gate, the ions could be exposed to a pulse of light from an optical parametric oscillator (OPO), tunable over the 300-700\,nm range. The parent ions and photoisomer ions were separated over the second drift region according to their collision cross-sections with \ce{N2} buffer gas. At the end of the drift region the ions were collected radially by a second ion funnel before passing through a 0.3\,mm orifice into an octopole ion guide housed in a differentially pumped vacuum chamber (P$\approx$5$\times$10$^{-4}$\,Torr), after which they passed through a quadrupole mass filter (P$\approx$5$\times$10$^{-6}$\,Torr) to a channeltron ion detector.  An arrival time distribution (ATD) for the ions was generated using a mutichannel scaler triggered at the same time as the first electrostatic ion gate. Normally, the ATD exhibits Gaussian peaks that correspond to the constituent isomers of the injected ion packet.  The mobility resolution of the instrument is $\frac{t_a}{\Delta t_a}$$\approx$80 (where $t_a$ and $\Delta t_a$ are arrival time and temporal width of the ion packet). The mass resolution of the quadrupole mass filter is $\Delta m$$\approx$3.
 
Photodetached electrons were detected by introducing trace \ce{SF6}, which has a large cross-section for capturing low energy electrons, into the \ce{N2} buffer gas in the drift region. The \ce{SF6-} anions (\mz~146) were temporally separated from the parent anions in the second drift region. 

\section{Computational approach}
Electronic structure calculations were performed using the Gaussian\,16, ORCA\,4.0.1  and CFOUR software packages.\cite{g16,orca,CFOUR} Geometrical optimizations and vibrational frequencies were computed at the CAM-B3LYP/aug-cc-pVDZ level of theory,\cite{CAMB3LYP,accd} followed by single-point energy calculations at the DLPNO-CCSD(T)/aug-cc-pVDZ level of theory.\cite{dlpno-cc} Vertical excitation energies for NPO$^{-}$ were computed at the EOM-CC3/aug-cc-pVDZ level of theory (excluding virtual orbitals with energies $>$\,2\,Hartree from the correlation space).\cite{eomcc3} Details of Franck-Condon-Herzberg-Teller (FCHT) modelling of the electronic transitions and vibronic structure are provided in the ESI.

Collision cross-sections were calculated using MOBCAL with the trajectory method parametrized for \ce{N2} buffer gas.\cite{Campuzano:2012jn,Mesleh:1996vf} Input charge distributions were computed at the CAM-B3LYP/aug-cc-pVDZ level of theory with the Merz-Singh-Kollman scheme constrained to reproduce the electric dipole moment.\cite{mk-pop} A sufficient number of trajectories was computed to give standard deviations of $\pm$1\,\AA{}$^{2}$ for the calculated values.

\section{Results and Discussion}
\subsection{Photodetachment of 2-naphtholate}

\begin{figure}
 \centering
 \includegraphics[width=7.5cm]{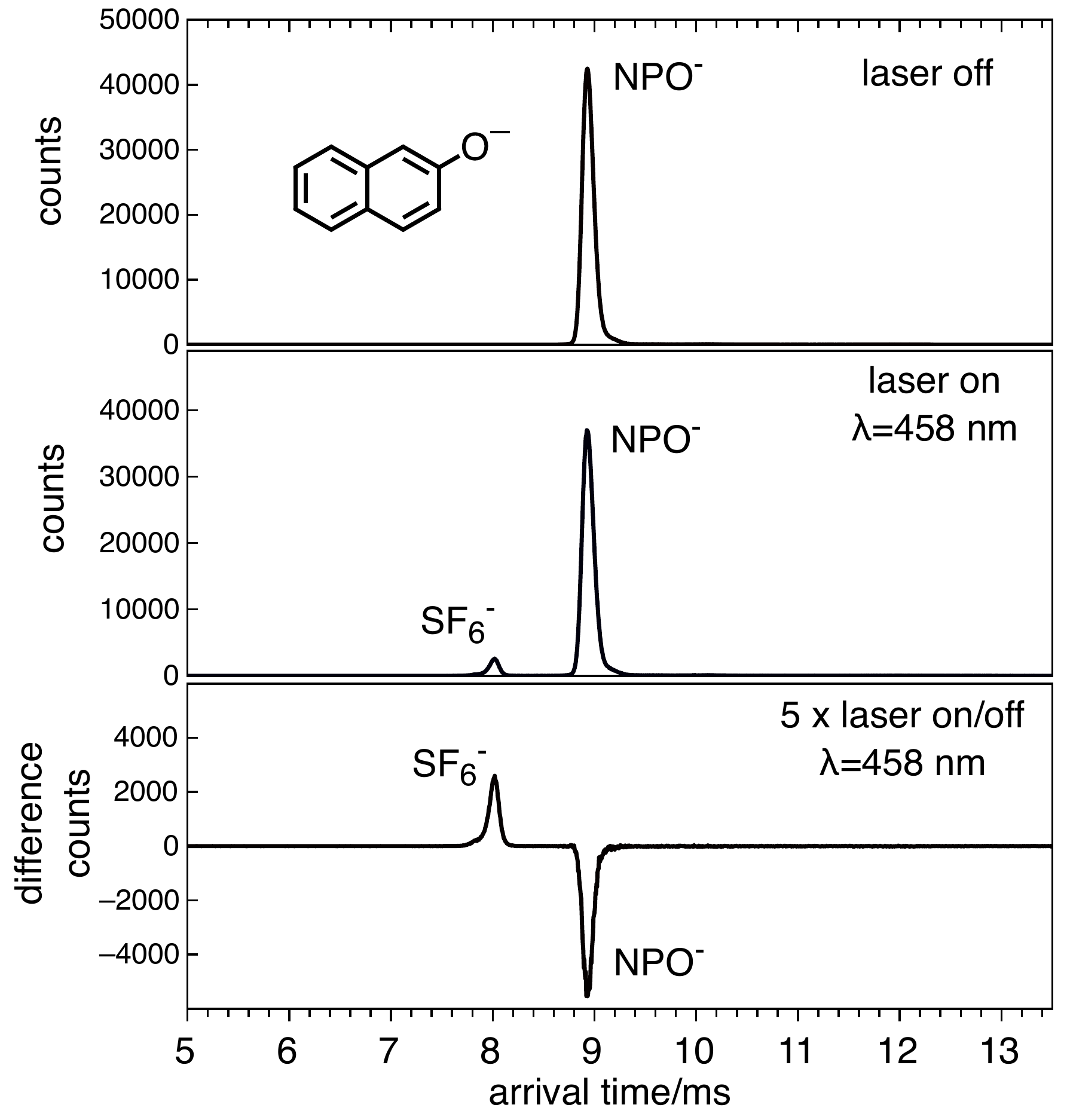}
 \caption{Arrival time distribution for 2-naphtholate anion (NPO$^-$) in \ce{N2} buffer gas (P$\approx$6\,Torr) with trace \ce{SF6}, which scavenges photodetached electrons. To record these ATDs the quadrupole mass filter was set to transmit all ions irrespective of mass. Depletion of the 2-naphtholate anion is apparently not balanced by creation of \ce{SF6-} due to mass-dependent transmission efficiency through IF2 and the quadrupole mass filter.}
  \label{f3}
\end{figure}

\begin{figure}
 \centering
 \includegraphics[width=8.cm]{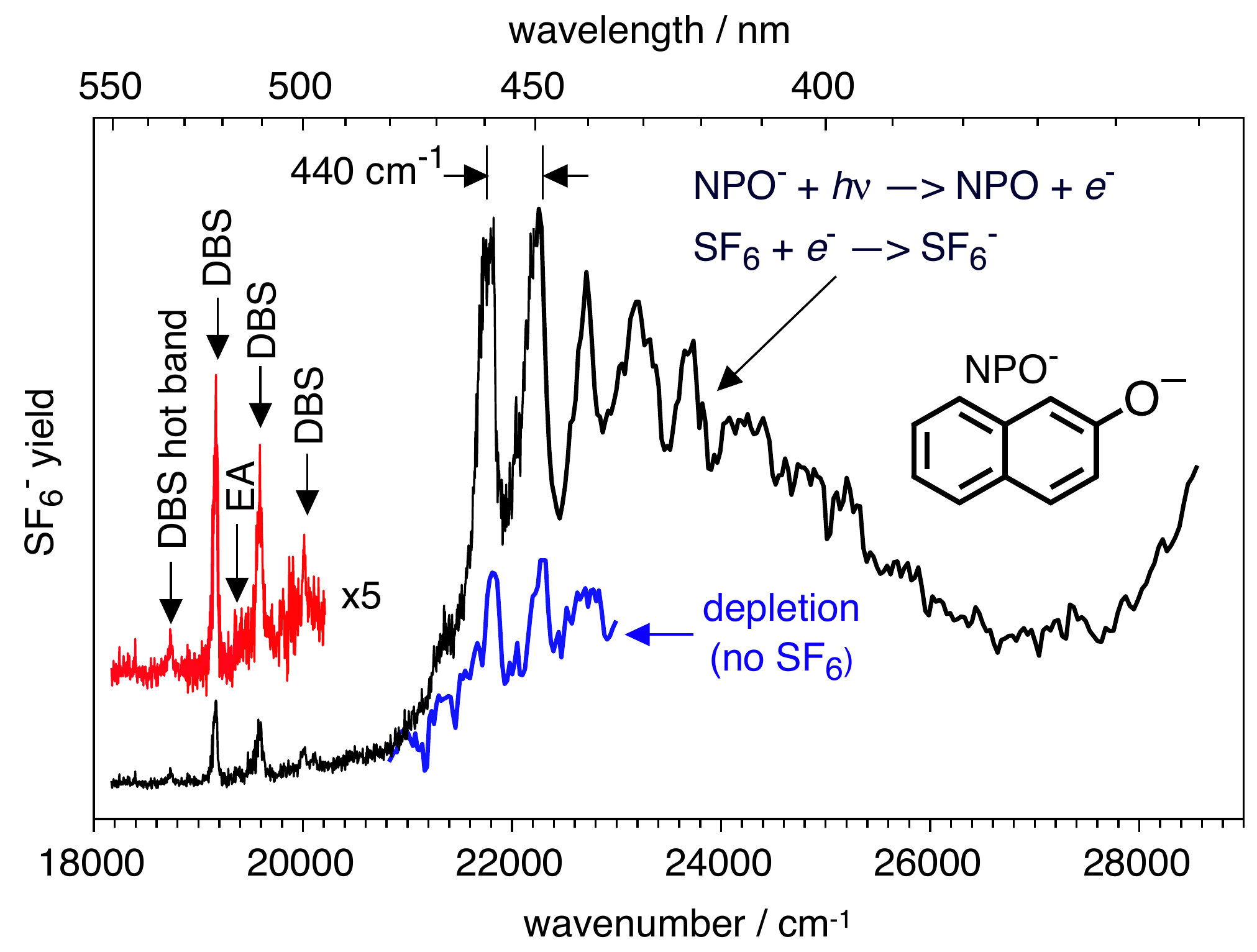}
 \caption{Photodetachment action spectrum of  2-naphtholate anion recorded by monitoring  \ce{SF6-} production as a function of excitation wavelength (black curves) or depletion of the 2-naphtholate anion in the absence of \ce{SF6} (blue curve). The proposed origin of the 2-naphtholate anion \ST~transition occurs at 21\,820\,\cme. Peaks corresponding to transitions to dipole-bounds states (DBS) below 21\,000\,\cme~are indicated with arrows. Also indicated is the electron detachment threshold (EA) at 19\,388\,\cme~(ref. \citenum{Kregel:2018dc}).}
  \label{f4}
\end{figure}

We first present and discuss the photodetachment spectrum of mobility-selected 2-naphtholate anions, illustrating one way in which electronic spectra of molecular anions can be obtained using the tandem IMS. In this case, the ion population generated by electrospraying a solution of 2-naphthol in methanol, when injected into the drift region, leads to a single peak in the arrival time distribution corresponding to the 2-naphtholate anion (see Fig.\,\ref{f3}). Selecting the 2-naphtholate anions by the Bradbury-Nielsen ion gate  situated after the first drift region and exposing them to light over the 400-500\,nm range resulted in photodepletion, presumably through photodetachment (electron affinity, EA=19\,388\,\cme; ref. \citenum{Kregel:2018dc}). Detached electrons were efficiently captured by \ce{SF6}, giving rise to a characteristic \ce{SF6-} peak in the ATD (see Fig.\,\ref{f3}). 

A photodetachment action spectrum was generated by monitoring the \ce{SF6-} peak as a function of wavelength. The spectrum, shown in Fig.\,\ref{f4}, has an onset at 21\,390\,\cme (467.5\,nm) and features a progression of bands spaced by $\approx$440\,\cme, the first of which occurs at 21\,820\,\cme~($\lambda$=458.5\,nm). 
Note, that the progression was also observed without \ce{SF6} in the drift tube through wavelength dependent photodepletion of the 2-naphtholate anion signal (blue curve in Fig.\,\ref{f4}), eliminating the possibility that the absorptions are due to a complex of the 2-naphtholate anion and \ce{SF6}. The observed band system is assigned to a \mbox{$\pi$-$\pi^*$} excitation with the upper electronic state lying above the electron detachment threshold (2.404\,eV -- 19\,388\,\cme, ref.\,\citenum{Kregel:2018dc}). The detachment mechanism is unknown but could involve either internal conversion to the ground S$_0$ state of the anion followed by vibrational autodetachment, or coupling of the S$_1$ state directly to the detachment continuum. The origin of the 2-naphtholate \ST~transition in the gas phase at 458.5\,nm lies around 8\,300\,\cme~lower than the \ST~transition in methanol solution ($\lambda$$\approx$332\,nm). The large shift is comparable to the 5\,000\,\cme~red shift for the phenolate anion, for which the lowest energy transition observed through photodetachment in the gas phase occurs at 330\,nm ($\approx$30\,000\,\cme),\cite{Richardson:2008js} compared to 285\,nm ($\approx$35\,000\,\cme) in aqueous solution. 

The electronic transitions of the 2-naphtholate anion should resemble those of the naphthalene molecule for which transitions to L$_a$ and L$_b$ states occur with the transition dipole moment aligned along the short and long axes of the molecule, respectively.\cite{Yang:2016} As outlined in the ESI, the L$_{a}$$\leftarrow$S$_0$ transition of 2-naphtholate is expected to occur near the observed band system, with EOM-CC3/aug-cc-pVDZ calculations predicting the vertical transition at $\lambda$=454\,nm with an oscillator strength of 0.14. Furthermore, simulations of the spectrum based on CAM-B3LYP/aug-cc-pVDZ calculations predict a progression with a spacing of 437\,\cme, corresponding to an in-plane ring deformation mode. The L$_{b}$$\leftarrow$S$_0$ transition is predicted to lie at around $\lambda$=423\,nm, and may be responsible for the spectrally unresolved signal in this region. More details of the calculations and comparisons of the measured and calculated spectra are given in the ESI.

We also considered whether the observed 2-naphtholate spectrum is associated with a triplet-triplet transition of the  anion, which in aqueous solution occurs at 460\,nm.\cite{Jackson:1961et} We first investigated whether triplet 2-naphtholate anions were formed through intersystem crossing (ISC) following excitation by residual 355\,nm in the OPO beam. However, the spectrum was still observed when 355\,nm light was eliminated using a dichroic reflector.   Alternatively, one could suppose that triplet 2-naphtholate anions were formed in the electrospray ion source making their way through the transfer capillary and the first stage of the drift tube to where they were intercepted by the tunable OPO beam. This would require that the  triplet anions' lifetime exceeded several hundred milliseconds (estimated time to pass through the transfer capillary and first section of the drift region), they have a very similar mobility to singlet 2-naphtholate anions (only a single ATD peak was observed), and they survive energetic collisions in the first ion funnel before injection into the drift region.

The photodetachment action spectrum shown in Fig.\,\ref{f4} exhibits several much weaker transitions at photon energies near the electron detachment threshold. The peaks above the detachment threshold at 19\,600$\pm$10 and 20\,020$\pm$20\,\cme~are associated with transitions to autodetaching, dipole-bound states (DBSs) as observed in slow electron velocity-map imaging spectroscopy of cryogenically cooled 2-naphtholate anions.\cite{Kregel:2018dc} An even stronger peak  below the detachment threshold at 19\,180$\pm$10\,\cme~likely arises from excitation of the lowest-lying DBS, which does not autodetach, but which is detected in the drift tube environment because the excited 2-naphtholate anion transfers an electron directly  to \ce{SF6} in a collisional encounter.\cite{Dunning:2005bw}  The spacing between the first two DBS peaks is $\approx$420\,\cme, presumably corresponding to an in-plane ring deformation vibrational mode. A much weaker peak at 18 730$\pm$10\,\cme, lying $\approx$450\,\cme~to lower energy from the origin, is most probably a hot band.  Importantly, observation of the recognised transitions to DBSs above the detachment threshold confirms the identity of the selected naphtholate anions.

\subsection{Photodetachment and photochemistry of 6-hydroxy-2-naphthoic acid deprotomers}

\begin{figure}
 \centering
 \includegraphics[width=7.5cm]{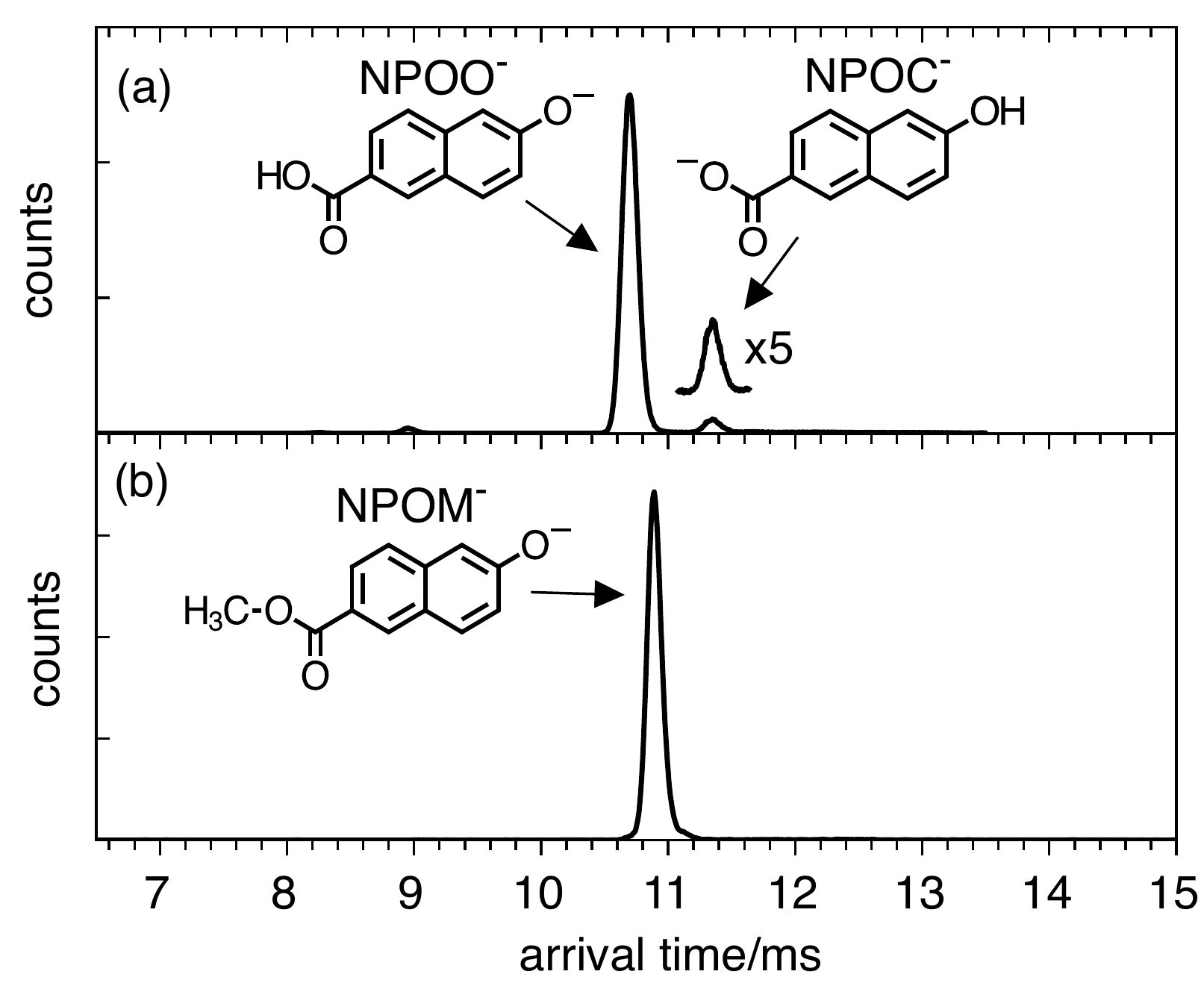}
 \caption{(a) Arrival time distribution for deprotonated 6-hydroxy-2-naphthoic acid in \ce{N2} buffer gas. The earlier peak at $\approx$11\,ms, corresponds to the NPOO$^-$ deprotomer and the later peak to the NPOC$^-$ deprotomer. The minor peak at $\approx$9\,ms is assigned to the doubly deprotonated dianion that loses an electron in IF2. (b) Arrival time distribution for the methyl ester of the 6-carboxy-2-naphtholate anion  (NPOM$^-$) in \ce{N2} buffer gas.}
 \label{f5}
\end{figure}

Deprotonated 6-hydroxy-2-naphthoic acid represents an interesting target as it can be deprotonated at  two sites to produce either the naphthoxide deprotomer (\npoo~- Fig.\,\ref{f1}(b)) or carboxylate deprotomer (\npoc~-- Fig.\,\ref{f1}(c)). In the gas phase, according to DLPNO-CCSD(T)/aug-cc-pVDZ level calculations,  the \npoo~naphthoxide deprotomer is more stable by 22\,kJ/mol than the \npoc~carboxylate deprotomer. However, in solution, the carboxylate form is favoured, based on pKa values for the two deprotonation sites. The ATD for electrosprayed 6-hydroxy-2-naphthoic acid exhibits two peaks (Fig.\,\ref{f5}(a)). On the basis of the calculated collision cross-sections ($\Omega_c$), the earlier peak is associated with the \npoo~isomer (Fig.\,\ref{f1}(c) $\Omega_c$=141\,\AA$^2$), whereas the later, weaker peak is associated with the naphthoate isomer (Fig.\,\ref{f1}(d)  $\Omega_c$=149\,\AA$^2$). 

\begin{figure}
 \centering
 \includegraphics[width=7.5cm]{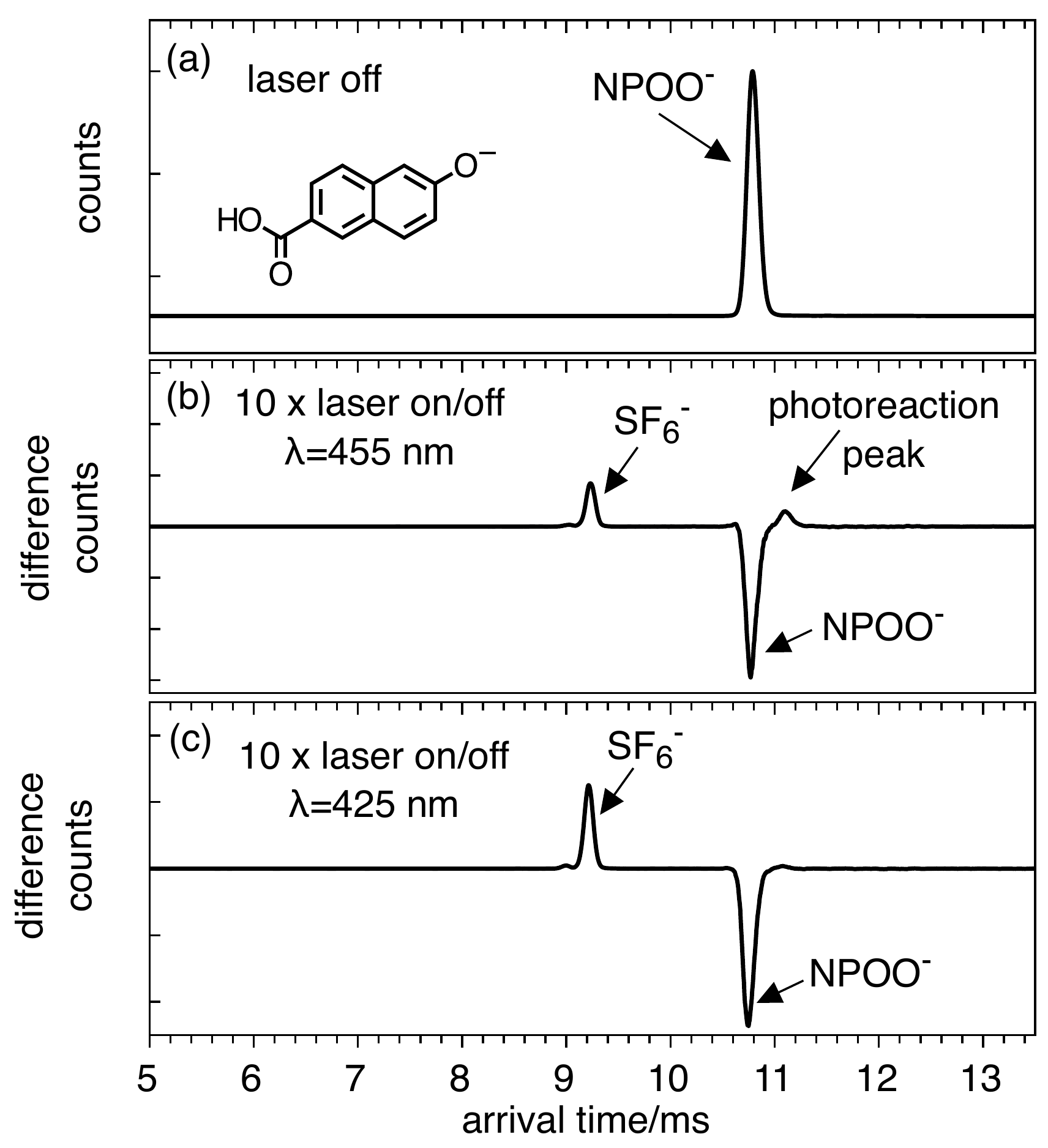}
 \caption{Photoaction ATD of the \npoo~anion in \ce{N2} (P=6\,Torr) buffer gas with trace \ce{SF6} recorded at 455\,nm (middle panel) and 425\,nm (lower panel). The small photoreaction peak in (b) corresponds to exchange of an O atom for an F atom. }
 \label{f6}
\end{figure}

\begin{figure}
 \centering
 \includegraphics[width=8.5cm]{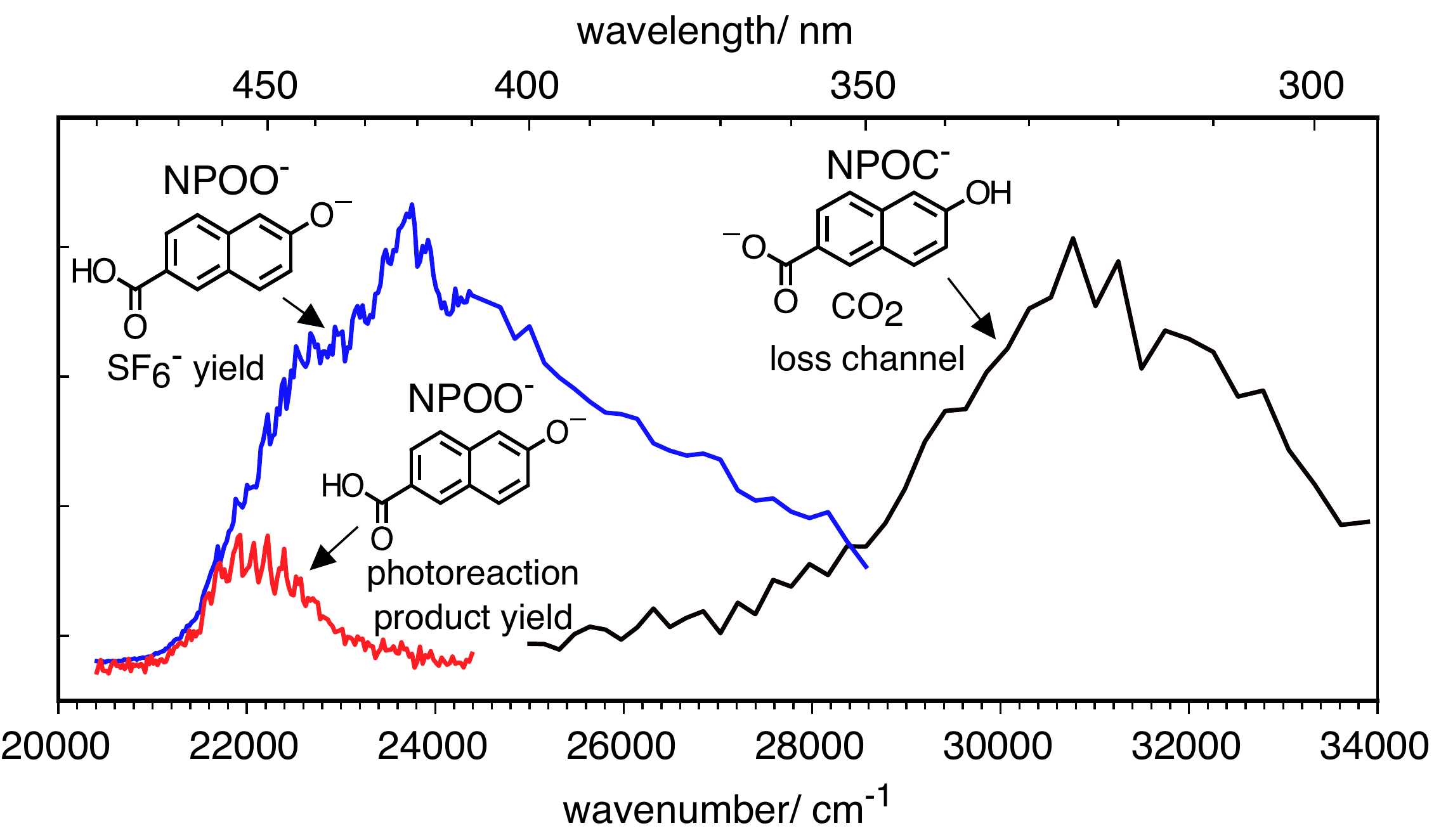}
 \caption{Electronic spectra of deprotonated 6-hydroxy-2-naphthoic acid anions. Shown are action spectra of the \npoo~deprotomer recorded by monitoring \ce{SF6-} production as a function of excitation wavelength (blue curve) and the $m/z$ 189 photoreaction product indicated in Fig.\,\ref{f5} (red curve). Also shown is the action spectrum of the  \npoc~deprotomer recorded by monitoring the \ce{CO2} loss channel.}
 \label{f7}
\end{figure}

The photochemical responses of the \npoo~and \npoc~isomers were investigated independently using the IMS-laser-IMS strategy.  Considering the \npoo~isomer first, we note that exposure to light below $\lambda$=460\,nm leads to photodetachment, where again photoelectrons were  captured by trace \ce{SF6}. The photoresponse is apparent in the action ATDs shown in Fig.\,\ref{f6}. The photodetachment action spectrum of the \npoo~isomer, recorded by measuring the \ce{SF6-} yield as a function of laser wavelength is shown in Fig.\,\ref{f7}. The spectrum has an onset at $\lambda$$\approx$470\,nm, with a peak response at 420\,nm, and resembles the photodetachment action spectrum of the 2-naphtholate anion albeit without the resolved vibronic structure. 

Intriguingly, some fraction of the excited \npoo~produce a slightly slower and heavier photoproduct ($m/z$ $\approx$189) (see Fig.\,\ref{f6}). The photoproduct is only generated in the presence of \ce{SF6} and only over a restricted wavelength range (470-430\,nm, see Fig.\,\ref{f7}). The most likely explanation is that the carboxyl OH group is displaced by an F atom following electronic excitation of \npoo. This explanation is consistent with measurements on NPOM$^{-}$ (i.e. the methyl ester derivative of \npoo, Fig.\,\ref{f1}(b)), which produced a similar  photodetachment action spectrum recorded by monitoring \ce{SF6-} (see ESI), but showed no evidence for the photoreaction. Similar photo-initiated deoxyfluorination reactions involving \ce{SF6} and a range of molecules have been observed in solution, although the mechanistic details are somewhat obscure.\cite{Rueping:2017hk,McTeague:2016ej, Rombach:2018gp,Tomar:2018jh} We investigated the formation of the photoproduct from \npoo~as a function of the \ce{SF6} partial pressure, which is conveniently assessed from the arrival time of the \npoo~peak; following Blanc's law,\cite{Blanc:1908oq} the increase in the arrival time should be proportional to the \ce{SF6} partial pressure.  The relative yields of the photoproduct and \ce{SF6-} are plotted in Fig.\,S6 in the ESI. The relative yield [photoproduct]/[\ce{SF6-}] approaches 0 at low \ce{SF6} partial pressure and reaches an asymptote of $\approx$0.5 as the \ce{SF6} pressure increases. This behaviour is consistent with efficient collection of detached electrons by \ce{SF6} (electrons are eventually captured even at low \ce{SF6} pressure) with photoproduct formation requiring a collisional encounter between \ce{SF6} and an electronically excited \npoo molecule. Given that the rate for \ce{SF6} and \npoo~collisions is 10$^5$--10$^6$\,s$^{-1}$  (assuming a \ce{SF6} partial pressure of 0.1\,Torr, and collision cross-section of $\approx$300\,\AA$^2$), one concludes that the  reacting \npoo~anions are in a long-lived excited electronic state (either a triplet state or DBS) accessed in a non-radiative transition from the S$_1$ state, remembering again that the reaction requires photo-excitation. Some evidence for involvement of a triplet state is that if the \ce{N2} buffer gas and trace \ce{SF6} was replaced by air and trace \ce{SF6}, the photoproduct channel was suppressed, conceivably because triplet \npoo~is quenched by \ce{O2} molecules. 

As described in the ESI, the absorptions of \npoo~in the visible region most likely arise from overlapping L$_{a}\leftarrow$S$_{0}$ and L$_{b}\leftarrow$S$_{0}$ transitions. Interestingly, the action spectrum for \npoo~recorded on the photoreaction product channel corresponds to the L$_{a}\leftarrow$S$_{0}$ transition and exhibits vibrational fine structure with a spacing of $\approx$160\,\cme, corresponding to an in-plane \ce{CO2} wag, as predicted by the FCHT calculations. The major band, centred at 425\,nm, observed on the photodetachment channel (\ce{SF6-} yield), appears to correspond to the L$_{b}\leftarrow$S$_{0}$ transition (see ESI for more information). TD-DFT calculations suggest the \npoo~deprotomer has other bright $\pi\pi^{*}$ states lying to higher energy that probably contribute to the short wavelength region of the band.

At this stage it is not clear why the $m/z$ 189 photoproduct ions are observed near the threshold for the L$_{a}\leftarrow$S$_{0}$ band but disappear as the excitation energy increases. 
One explanation might be that generation of the photoproduct involves intersystem crossing and formation of triplet \npoo, but that with increasing excitation energy the rate for electron detachment from these triplet \npoo~anions increases and begins to exceed  the rate for reactive encounters with \ce{SF6} molecules so that the triplet \npoo~anions are lost before they can react.

The photodissociation action spectrum of the \npoc~isomer, recorded on the \ce{CO2} loss channel differs markedly from the \npoo spectrum (see Fig.\,\ref{f7}), with the appearance of a broad band with an onset at 360\,nm and a maximum at 325\,nm. Note that it was not possible to record a photodetachment action spectrum for \npoc~because of interference from background photoelectrons emanating from metal surfaces at wavelengths below $\lambda$=350\,nm.  The substantially shorter wavelength for the  L$_{a}\leftarrow$S$_{0}$ transition of \npoc~compared to that of \npoo~is consistent with calculations at the EOM-CC2/aug-cc-pVDZ level of theory which predict vertical transition energies for NPOO$^{-}$ and NPOC$^{-}$ of 2.98\,eV ($\lambda$=416\,nm) and 3.97\,eV ($\lambda$=312\,nm), respectively (see ESI for details).  Comparison of the \npoo and \npoc action spectra in Fig.\,\ref{f7} highlights the importance of using an IMS stage to separate the two isobaric deprotomers prior to spectroscopic interrogation.

\section{Conclusions and Outlook}
In this paper we show that the tandem IMS-laser-IMS approach  can be deployed to investigate substituted naphthalene anions that undergo photodissociation or photodetachment (with the photodetached electrons conveniently captured by \ce{SF6}). In the case of deprotonated 6-hydroxy-2-naphthoic acid the two deprotomers are easily separated and are found to have distinct electronic spectra. Surprisingly, for anions in a drift tube environment, it was possible to not only measure resonance enhanced photodetachment action spectra associated with valence transitions, but also to observe transitions to dipole-bound states. The 2-naphtholate anion displays a resolved vibrational progression extending down from 458.5\,nm, with a spacing of 440\,\cme. It will interesting to see whether spectra recorded for cooled 2-naphtholate ions display narrower band and additional spectral features.

Technical advances should extend the array of molecular ions that can be studied using drift tube techniques and improve the quality of the spectra. First, photodetachment action spectra recorded using current instruments are relatively broad and normally do not exhibit resolved vibronic structure. The situation may be improved by using a cryogenically cooled drift tube in order to suppress spectral congestion associated with hot bands and broad rotational profiles. Second, in future there is likely to be greater application of approaches in which molecular ions are selected using an IMS stage prior to trapping and probing in a cryogenic ion trap. The advantages are particularly relevant for studies of carbon and metal cluster systems for which there are often several different co-existing structures (with distinct electronic and reactive properties) that are not easy to separate and isolate using mass spectrometers alone.

\section*{Conflicts of interest}
There are no conflicts to declare.

\section*{Acknowledgements}
This research was supported under the Australian Research Council's Discovery Project funding scheme DP150101427 and DP160100474, by the Swedish Research Council (grant number 2016-03675) and by the Swedish Foundation for International Collaboration in Research and Higher Education (grant number PT2017-7328).

\balance



\providecommand*{\mcitethebibliography}{\thebibliography}
\csname @ifundefined\endcsname{endmcitethebibliography}
{\let\endmcitethebibliography\endthebibliography}{}

\end{document}